\documentclass[amsmath,11pt]{article}

\date{\empty}

\begin{document}

\title{\bf On constraint-consistency, covariant operators,
gauge-invariance, etc}

\author{Christos G. Tsagas\\ {\small Section of Astrophysics,
Astronomy and Mechanics, Department of Physics}\\ {\small Aristotle
University of Thessaloniki, Thessaloniki 54124, Greece}}

\maketitle

\begin{abstract}
We look at the covariant techniques and the ideas on constraints and
gauge-invariance, which were recently employed in~\cite{BZDM2} to
support earlier work by the same authors. That work was criticised
in~\cite{T2}. Using very simple and well known examples we show
that, when adopted, the methods and views of~\cite{BZDM2} lead to
basic-level mathematical problems, with analogous consequences for
the physics. We provide a few simple rules that should help to avoid
similar problems in the future.
\end{abstract}

\section{Introduction}
The subject of this note is related to the 1+3-covariant study of
the interaction between gravitational waves and large-scale magnetic
fields in cosmology. Our aim is not to discuss this interaction
again. The literature is available to anyone who might be
interested. Here, we will look at the mathematics (as recently
explained in~\cite{BZDM2}) on which the physics of~\cite{BZDM1} was
based upon.

The main issues have to do with the consistency of constraints, the
use of 3-D commutation operators and gauge-invariance. The problems
are technical but also elementary. The authors imposed inconsistent
constraints, used incomplete commutation laws and presented as
gauge-invariant results that are gauge-dependent by construction. We
argue mainly by using simple and well known examples and provide a
few basic rules that should help to avoid analogous problems in the
future. Given that, the present brief communique may also have some
pedagogical value.

\section{Constraints}
It was claimed (see \S~$III.B$ in~\cite{BZDM1}) that, despite the
presence of an inhomogeneous magnetic field, the electric component
stays curl-free, provided it was so initially and the zero-order FRW
model is spatially flat. The vanishing of ${\rm curl}E_a$ does not
depend on the scale, or the electrical conductivity, and applies to
the second perturbative order. Even without looking at the
mathematics, this claim sounds rather extreme.

A constraint is consistent if, once imposed, it holds at all times
with no need for further restrictions. To check whether ${\rm
curl}E_a$ vanishes, we must look for sources in the propagation
equation (i.e.~the first time-derivative) of ${\rm curl}E_a$.
In~\cite{BZDM1} the authors considered instead the second
time-derivative of this quantity, which on a spatially flat FRW
background reads (see Eq.~(29) in~\cite{BZDM1})
\begin{equation}
({\rm curl}E_a)^{\cdot\cdot}= -{7\over3}\,\Theta({\rm
curl}E_a)^{\cdot}+ {\rm D}^2{\rm curl}E_a-
\left[{7\over9}\,\Theta^2+{1\over6}\,(\rho-9p)
+{5\over3}\,\Lambda\right]{\rm curl}E_a\,. \label{eq:D1}
\end{equation}
Here, ${\rm curl}E_a=\varepsilon_{abc}{\rm D}^bE^c$, $\Theta$ is the
background volume expansion, $\rho$ and $p$ are the matter density
and pressure, $\Lambda$ is the cosmological constant, ${\rm D}_a$ is
the 3-D covariant derivative and ${\rm D}^2={\rm D}^a{\rm D}_a$ is
the associated Laplacian. The absence of explicit source terms in
Eq.~(\ref{eq:D1}) misled the authors into claiming that ${\rm
curl}E_a$ will remain zero, if it was so initially (see \S~$III.B$
in~\cite{BZDM1} and also~\cite{BZDM2}). This conclusion is
incorrect, as the following simple example shows.

Consider a perturbed FRW universe with dust. It is well known that
the second time-derivative of the density perturbation (here
represented by $\Delta$) reads
\begin{equation}
\ddot{\Delta}= -{2\over3}\,\Theta\dot{\Delta}+
{1\over2}\,\rho\Delta\,. \label{ddotD}
\end{equation}
Just like Eq.~(\ref{eq:D1}), the above contains no explicit source
terms. Then, if we were to adopt~\cite{BZDM2,BZDM1}, $\Delta$ will
stay zero if it was initially zero and there will be no density
perturbations.\footnote{Initially, refers to the initial perturbed
hypersurface and not to the FRW background, as~\cite{BZDM2} seems to
suggest.} We all know that this is not the case. To look for sources
one must check the first time-derivative of $\Delta$, namely the
expression
\begin{equation}
\dot{\Delta}= -\mathcal{Z}\,.  \label{dotD}
\end{equation}
This ensures that density perturbations are sourced by those in the
expansion (represented here by $\mathcal{Z}$). For the same reasons,
when dealing with ${\rm curl}E_a$, one must look for sources in the
first time-derivative of this quantity. The latter gives (see
Eq.~(1) in~\cite{T2})
\begin{equation}
({\rm curl}E_a^{(2)})^{\cdot}= -\Theta{\rm curl}E_a^{(2)}+
\mathcal{R}_{ab}^{(1)}B_{(1)}^b-{\rm D}^2B_a^{(2)} - {\rm
curl}\mathcal{J}_a^{(2)}\,, \label{1}
\end{equation}
where $\mathcal{R}_{ab}$ is the 3-Ricci tensor, $\mathcal{J}_a$ is
the 3-current and the indices $(1)$, $(2)$ indicate the perturbative
order of the variable. Thus, even if ${\rm curl}E_a=0$ initially, it
will not remain so once the magnetic and the current perturbations
kick in (unless further constraints -- on the sources -- are
imposed). The zero electric-curl claim made in~\cite{BZDM1} is
unsustainable.

Note that, in line with~\cite{BZDM1}, the $B$-field vanishes to zero
order has a homogeneous first-order component ($B_a^{((1)}$) and an
inhomogeneous second-order part ($B_a^{(2)}$). Thus,
$B_a=B_a^{((1)}+B_a^{(2)}$. Splitting the variables like that is
largely avoided in covariant calculations, but we will adopt it here
to agree with~\cite{BZDM2}.

\section{Commutators}\label{sCs}
The 3-D commutators have the form $2{\rm D}_{[a}{\rm D}_{b]}={\rm
D}_a{\rm D}_b-{\rm D}_b{\rm D}_a$ and monitor the commutation of 3-D
covariant derivatives. How to use these operators became an issue,
when~\cite{BZDM2} sought to remove the source terms from the
right-hand side of Eq.~(\ref{1}). One can imagine situations where
some of these sources are negligible relative to the rest, but not
all three of them simultaneously. When the Laplacian is the only
quantity left in the right-hand side of~(\ref{1}), in particular, we
cannot ignore it (even on large -- finite-- scales) because it
vanishes only asymptotically (i.e.~at infinity). Here, we will focus
on the curvature term because the argument used to remove this
quantity suffers at the elementary level.

The curvature term in Eq.~(\ref{1}) appears when analysing the
quantity ${\rm curl\,curl}B_a$ by means of the covariant 3-D
commutator ${\rm D}_{[a}{\rm D}_{b]}$. This operator applies to the
full variable and the commutation takes place prior to any
decomposition (into background, first-order perturbation, etc -- the
reasons will become clear below). It is argued in~\cite{BZDM2} (see
Eq.~(3) there), that we must instead split the variables first and
commute their 3-D gradients afterwards. Splitting the magnetic field
into $B_a=B_a^{(1)}+B_a^{(2)}$, the authors involve only the
inhomogeneous $B_a^{(2)}$-part and arrive at the second-order
formula
\begin{equation}
{\rm curl\,curl}B_a= {\rm curl\,curl}B_a^{(2)}= -{\rm D}^2B_a^{(2)}+
\mathcal{R}_{ab}^{(0)}B_{(2)}^b= -{\rm D}^2B_a^{(2)}\,, \label{2}
\end{equation}
since $\mathcal{R}_{ab}^{(0)}=0$ (see Eq.~(3) in~\cite{BZDM2}). The
term $\mathcal{R}_{ab}^{(1)}B^b_{(1)}$ of (\ref{1}) does not appear
above, although it is also second order and it should have been
included.

We will accept, for argument's sake, the recipe leading to
expression (\ref{2}) and test it on a well known case. Assume a
perturbed FRW universe containing a single perfect fluid with
nonzero pressure. Following~\cite{BZDM2}, we decompose the density
as $\rho=\rho^{(0)}+\rho^{(1)}$, where $\rho^{(0)}$ is the
homogeneous (zero-order) part and $\rho^{(1)}$ the inhomogeneous
perturbation. Then, in line with~\cite{BZDM2} and Eq.~(\ref{2}),
only $\rho^{(1)}$ takes part in the operation and the linear
commutator for the 3-gradients of $\rho$ reads
\begin{equation}
{\rm D}_{[a}{\rm D}_{b]}\rho= {\rm D}_{[a}{\rm D}_{b]}\rho^{(1)}=
-\dot{\rho}^{(1)}\omega_{ab}^{(0)}=0\,,  \label{3}
\end{equation}
since $\omega_{ab}^{(0)}=0$. However, almost everyone using the
covariant equations knows that the appropriate linear expression is
\begin{equation}
{\rm D}_{[a}{\rm D}_{b]}\rho= {\rm D}_{[a}{\rm D}_{b]}\rho^{(1)}=
-\dot{\rho}^{(0)}\omega_{ab}^{(1)}\neq0\,, \label{4}
\end{equation}
obtained easily when we commute first and split
afterwards.\footnote{The reader might wonder why it should make a
difference which operation (splitting or commuting) is done first.
It should not in principle, but if we first split and then commute
the chances of doing something wrong increase considerably.} The
commutation recipe employed in~\cite{BZDM2} is incomplete. Note that
if we were to use relation (\ref{3}) instead of (\ref{4}), the
linear propagation equation of the vorticity would be given by
\begin{equation}
\dot{\omega}_{ab}=-{2\over3}\,\Theta\omega_{ab}\,,  \label{5}
\end{equation}
instead of the familiar expression
\begin{equation}
\dot{\omega}_{ab}=
-{2\over3}\left(1-{3\over2}\,c_s^2\right)\Theta\omega_{ab}\,.
\label{6}
\end{equation}
The extra term in the parentheses results from the right-hand side
of Eq.~(\ref{4}) -- $c_s^2$ is the square of the adiabatic
sound-speed. Similar problems will emerge in several other
expressions, if we were to follow the commutation rule
of~\cite{BZDM2}, and relativistic cosmology would require drastic
revision.

\section{Gauge-invariance}
An additional issue is that of gauge-invariance.
In~\cite{BZDM2,BZDM1} the authors consider the aforementioned
$B$-field, which has nonzero first-order value and is therefore
gauge-dependent at second order by well known theorems. It is noted
that the first-order field has
\begin{equation}
\dot{B}_a^{(1)}+{2\over3}\,\Theta B_a^{(1)}=0\,, \label{7}
\end{equation}
which makes the quantity
\begin{equation}
\beta_a= \dot{B}_a+{2\over3}\,\Theta B_a=
\dot{B}_a^{(2)}+{2\over3}\,\Theta B_a^{(2)}\,,  \label{8}
\end{equation}
a gauge-independent variable at second order. This is true, although
there should be a $2\Theta^{(1)}B^{(1)}/3$ term somewhere in the
above (to simplify things lets say that $\Theta^{(1)}=0$). The real
problem starts when the authors claim that they can integrate
Eq.~(\ref{8}), with respect to $B_a^{(2)}$, and obtain
gauge-independent results (see \S~$IV$ in~\cite{BZDM2}). This is an
interesting idea, but it does not sound right. The gauge ambiguity
resides in $B_a^{(2)}$ itself, by construction, due to the nonzero
first-order value of $B_a$.

Let us assume, for argument's sake again, that~\cite{BZDM2} are
right and apply their principle to a familiar case. Consider a
perturbed FRW universe containing dust with density
$\rho=\rho^{(0)}+\rho^{(1)}$. This quantity is the archetype of a
gauge-dependent perturbation. To zero order,
\begin{equation}
\dot{\rho}^{(0)}+\Theta\rho^{(0)}=0\,,  \label{9}
\end{equation}
which means that the quantity
\begin{equation}
\varrho= \dot{\rho}+ \Theta\rho= \dot{\rho}^{(1)}+ \Theta\rho^{(1)}
\label{10}
\end{equation}
is a gauge-invariant first-order perturbation.
Following~\cite{BZDM2}, one can integrate the above and obtain
gauge-independent results for $\rho^{(1)}$. The same principle could
be applied to every gauge-dependent perturbation. If this were true,
there should be no need to introduce complicated variables to
describe gauge-invariant perturbations and the gauge problem would
have been a rather trivial one. Unfortunately, it is not that
simple. Nevertheless, if the authors still believe in their case,
they should communicate their findings without delay.

\section{Numerical results}
It is also worth mentioning the numerical evaluation of
Eq.~(\ref{11}) (see expression (7) in~\cite{BZDM2}). The latter
gives the strength of a gravitationally amplified $B$-mode that
passes through reheating and enters the horizon sometime in the
radiation era
\begin{equation}
B=\tilde{B}_0\left[1 +\left({a_{RH}\over a_o}\right)_0
\left({\sigma\over H}\right)_0
+\left({\lambda_{\tilde{B}}\over\lambda_H}\right)_{RH}^2
\left({\sigma\over H}\right)_{RH}
+\left({\lambda_{\tilde{B}}\over\lambda_H}\right)_0^2
\left({\sigma\over H}\right)_0 \left({\sigma\over
H}\right)_{RH}\right]\left({a_0\over a}\right)^2\,.  \label{11}
\end{equation}
Here $\sigma/H$ describes the shear anisotropy,
$\lambda_{\tilde{B}}$ and $\lambda_H$ are the scale of the $B$-field
and the horizon size, while the indices $0$ and $RH$ indicate
evaluation at the end of inflation and reheating respectively (with
$\tilde{B}=B^{(1)})$. The physics leading to the above result was
explained in~\cite{T2}. It was clarified there that the
amplification is of the superadiabatic nature, takes place outside
the horizon and defers from epoch to epoch. This view has also been
adopted in~\cite{BZDM2}, but the authors dropped the last term in
the brackets claiming that it is of the third order. This is not so,
as a bit of algebra can easily show.\footnote{It helps to recall
that the amplified $B$-field at the end of reheating, acts as the
background source for the gravito-magnetic interaction in the
radiation era (see~\cite{T2}).} What is more important, is that the
quantities in the brackets are numbers and what decides which stays
or goes is their value.

As explained in~\cite{T2} (see Eq.~(10) there), when there is
substantial magnetic amplification during both the reheating and the
radiation epochs, the third term in (\ref{11}) prevails. All these
are of uncertain practical value, however, as long as the
gauge-issue remains unresolved.

\section{Conclusions}
We close by summarising this communique in four basic rules: (i) To
check the consistency of a constraint we look at the first (not the
second) time-derivative of the constrained quantity; (ii) To avoid
mistakes, we first commute the 3-D gradients and then split the
involved variables (not the other way around); (iii) The
gauge-ambiguity and the arbitrariness are in the variable that
describes the perturbation; (iv) When evaluating sums like that of
Eq.~(\ref{11}), one should always ensure that all the terms are
properly accounted for.

\end{document}